\documentclass{llncs}

\usepackage[ansinew]{inputenc}
\usepackage[dvips]{graphicx}
\usepackage{amsmath}
\usepackage{amssymb}
\usepackage{enumerate,cite}
\usepackage{multirow}

\newcounter{algo}
\newdimen\ndec
\newcommand{\ns}[1]{\vskip 0.05cm\stepcounter{algo}\hskip 0cm\hbox to
\hsize{{\thealgo.} #1\hfill}\vskip 0cm}

\newcommand{\nl}[1]{\vskip 0.05cm\hangindent=\ndec \hangafter=0 %\noindent
\noindent\hsize=\lgcomment#1\vskip 0cm}

\newcommand{\np}[1]{\vskip 0.00cm \hskip 0cm\hbox to
\hsize{#1\hfill}\vskip -0.00cm}

\newcounter{noprocedure}%[section]

\newcounter{noalgo}%[section]
\renewcommand{\thenoalgo}{\arabic{noalgo}}
\def\algoname{Algorithm }
\newdimen\longueur

\newenvironment{algo}[3]{
\par
\bigskip
\small
\newdimen\decalage
\newdimen\lgcomment
\longueur=\textwidth

\hsize=\longueur
\lgcomment=\hsize
\advance\lgcomment by \decalage
\parindent=\decalage
\longueur=\hsize
\refstepcounter{noalgo} %TRES important : qd on met un label dans le corps de l'algo, c'est noalgo qui sera
\setcounter{algo}{0}
 \vbox{
 \hskip0cm
\raise 20pt\hbox{\hskip \decalage\hbox to \hsize{\hrulefill}}}
\vskip -0.8cm
\hangindent=\decalage
\hangafter=1
\hsize=\longueur
\advance\hsize by\decalage
{\bf \algoname \thenoalgo} \hbox to 0.1cm{}#1\par
\vskip -0.3cm
\hskip 0cm\hbox to \longueur{\hrulefill}
\vskip -0.1cm
\hsize=\longueur
\advance\hsize by\decalage
\hangindent=\decalage
\hangafter=1
{\sc Inputs :} #2\par
\hangindent=\decalage
\hangafter=1
\hsize=\longueur
\advance\hsize by\decalage
\vskip -0.1cm
{\sc Output :} #3\par
\vskip -0.3cm
\hskip 0cm\hbox to \longueur{\hrulefill}\par
\vskip 0.1cm
}
{
\vskip -0.1cm
\hsize=\longueur
\parindent=\decalage
\hskip 0cm\hbox to \hsize{\hrulefill}\vskip 0.3cm\par
}

\newcommand{\While}{{\bf while }}
\newcommand{\Ret}{{\bf return }}
\newcommand{\Do}{{\bf\hskip0.9mm do }}
\newcommand{\If}{{\bf if }}

\newcommand{\Else}{{\bf\hskip0.9mm else }}

\newcommand{\Then}{{\bf\hskip0.9mm then }}

\renewcommand{\O}{\mathcal{O}}
\newcommand{\F}{\mathbb F}

\newcommand{\E}{\mathcal{E}}

\newcommand{\EFp}{\mathcal{E}\left(\mathbb{F}_p\right)}

\newif\ifpdf
\pdftrue

\usepackage{url}
\usepackage{color}
\definecolor{rouge}{gray}{.7}

\title{Atomicity Improvement for Elliptic Curve Scalar Multiplication}

\author{Christophe Giraud\inst{1} \and Vincent Verneuil\inst{2,3}\thanks{A part of this work has been done while at Oberthur Technologies.}}
\institute{Oberthur Technologies,\\4, all\'ee du doyen Georges Brus, 33\,600 Pessac, France\\\email{c.giraud@oberthur.com}
\and Inside Contactless,\\41, parc du Golf, 13\,856 Aix-en-Provence cedex 3, France\\\email{vverneuil@insidefr.com}
\and Institut de Math\'ematiques de Bordeaux,\\351, cours de la Lib\'eration, 33\,405 Talence cedex, France}

\begin{document}

\maketitle

\begin{abstract}
In this paper we address the problem of protecting elliptic curve scalar multiplication implementations against side-channel analysis by using the atomicity principle. First of all we reexamine classical assumptions made by scalar multiplication designers and we point out that some of them are not relevant in the context of embedded devices. We then describe the state-of-the-art of atomic scalar multiplication and propose an atomic pattern improvement method. Compared to the most efficient atomic scalar multiplication published so far, our technique shows an average improvement of up to 10.6\%.
\end{abstract}

\section{Introduction}
\enlargethispage{10pt}

\subsection{Preamble}

We consider the problem of performing scalar multiplication on elliptic curves over $\F_p$ in the context of embedded devices such as smart cards. In this context, efficiency and side-channel resistance are of utmost importance. Concerning the achievement of the first requirement, numerous studies dealing with scalar multiplication efficiency have given rise to efficient algorithms including sliding-window and signed representation based methods~\cite{HMV04}.

Regarding the second requirement, side-channel attacks exploit the fact that physical leakages of a device (timing, power consumption, electromagnetic radiation, etc) depend on the operations performed and on the variables manipulated. These attacks can be divided into two groups: the \emph{Simple Side-Channel Analysis} (SSCA)~\cite{Koc96} which tries to observe a difference of behavior depending on the value of the secret key by using a single measurement, and the \emph{Differential Side-Channel Analysis} (DSCA)~\cite{KJJ99} which exploits data value leakages by performing statistical treatment over several hundreds of measurements to retrieve information on the secret key.
Since 1996, many proposals have been made to protect scalar multiplication against these attacks~\cite{Cor99,JT01,BJ02b}. Amongst them, \emph{atomicity} introduced by Chevallier-Mames et al. in~\cite{CCJ04} is one of the most interesting methods to counteract SSCA. This countermeasure has been widely studied and Longa recently proposed an improvement for some scalar multiplication algorithms~\cite{Lon07}.

In this paper we present a new atomicity implementation for scalar multiplication, and we detail the atomicity improvement method we employed. This method can be applied to minimize atomicity implementation cost for sensitive algorithms with no security loss. In particular our method allows the implementation of atomic scalar multiplication in embedded devices in a more efficient way than any of the previous methods.

The rest of this paper is organized as follows. We finish this introduction by describing the scalar multiplication context which we are interested in and by mentioning an important observation on the cost of field additions. In Section~\ref{sec:eca} we recall some basics about Elliptic Curves Cryptography. In particular we present an efficient scalar multiplication algorithm introduced by Joye in 2008~\cite{Joy08a}. Then we recall in Section~\ref{sec:atomicite} the principle of atomicity and we draw up a comparative chart of the efficiency of atomic scalar multiplication algorithms before this work. In Section~\ref{sec:atomic-improvement}, we propose an improvement of the original atomicity principle. In particular, we show that our method, applied to Joye's scalar multiplication, allows a substantial gain of time compared to the original atomicity principle. Finally, Section~\ref{sec:conclusion} concludes this paper.

\subsection{Context of the Study}

We restrict the context of this paper to practical applications on embedded devices which yields the constraint of using standardized curves over $\F_p$\footnote{The curves over $\F_p$ are generally recommended for practical applications~\cite{SP800781,TR03111}.}.
As far as we know, NIST curves~\cite{FIPS186-3} and Brainpool curves~\cite{Bra05,Bra09} cover almost all curves currently used in the industry. We thus exclude from our scope Montgomery curves~\cite{Mon87}, Hessian curves~\cite{Hes44}, and Edwards curves\footnote{An elliptic curve over $\F_p$ is expressible in Edwards form only if it has a point of order 4~\cite{BL07b} and is expressible in twisted Edwards form only if it has three points of order 2~\cite{BBJ08}. Since NIST and Brainpool curves have a cofactor of 1 there is not such equivalence. Nevertheless, for each of these curves, it is possible to find an extension field $\F_{p^q}$ over which the curve has a point of order 4 and is thus birationally equivalent to an Edwards curve. However the cost of a scalar multiplication over $\F_{p^q}$ is prohibitive in the context of embedded devices.}~\cite{Edw07} which do not cover NIST neither Brainpool curves.

Considering that embedded devices -- in particular smart cards -- have very constrained resources (i.e. RAM and CPU), methods requiring heavy scalar treatment are discarded as well. In particular it is impossible to store scalar precomputations for some protocols such as ECDSA~\cite{ANSIX9.62-2005} where the scalar is randomly generated before each scalar multiplication. Most of the recent advances in this field cannot thus be taken into account: Double Base Number System~\cite{DIM05,HM09}, multibase representation~\cite{LM08}, Euclidean addition chains and Zeckendorf representation~\cite{Mel07}.

\subsection{On the Cost of Field Additions}
\label{add_cost}

\begin{figure}[ht]%
\center
\ifpdf
\includegraphics[width=.65\columnwidth]{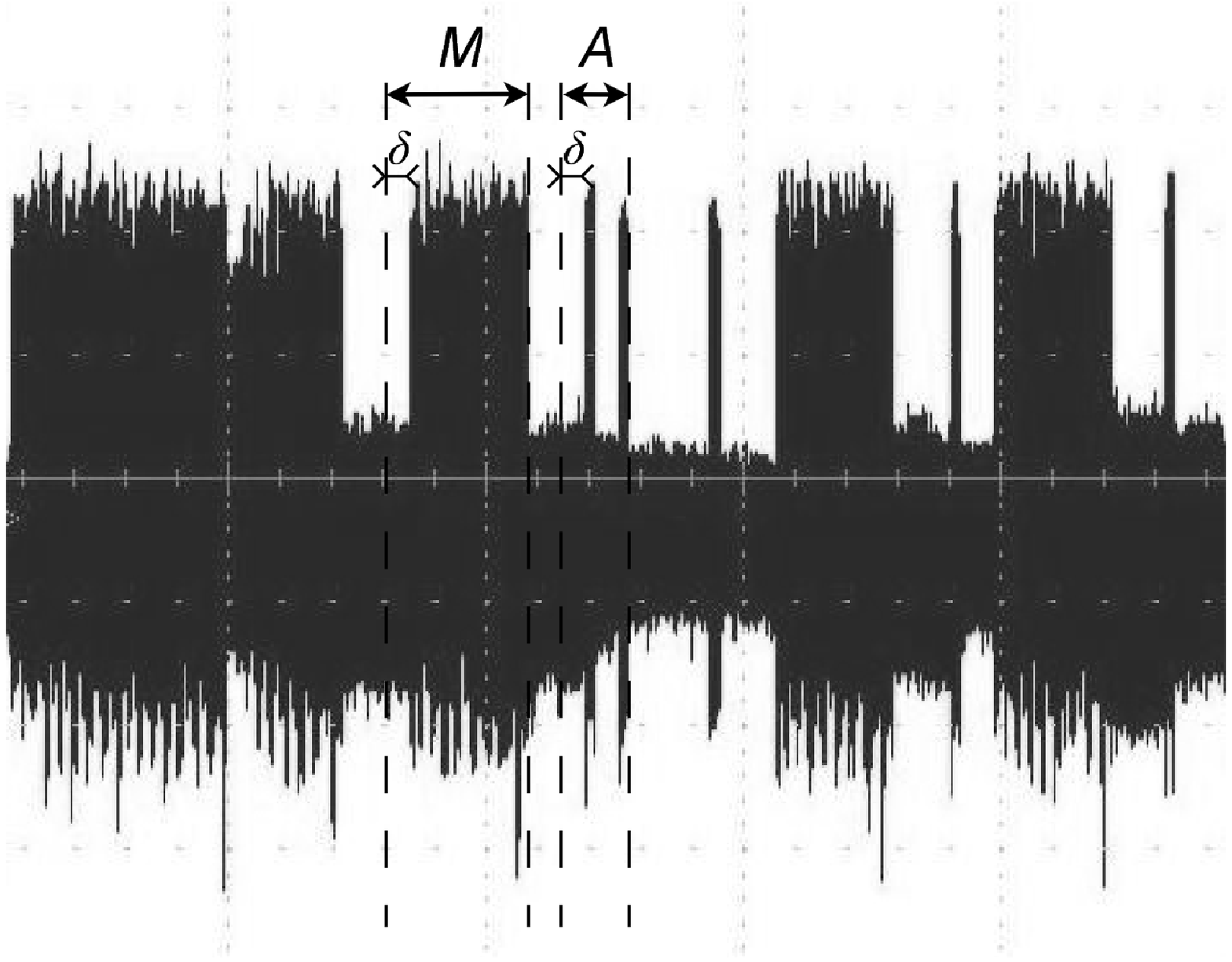}%
\else
\includegraphics[bb=0 0 1308 1028,width=.65\columnwidth]{EMA_1.eps}%
\fi 
\caption{Comparison between modular multiplication (M) and modular addition (A) timings.}%
\label{fig_ema}%
\end{figure}

In the literature, the cost of additions and subtractions over $\F_p$ is generally neglected compared to the cost of field multiplication. While this assumption is relevant in theory, we found out that these operations were not as insignificant as predicted for embedded devices. Smart cards for example have crypto-coprocessors in order to perform multi-precision arithmetic. These devices generally offer the following operations: addition, subtraction, multiplication, modular multiplication and sometimes modular squaring. Modular addition (respectively subtraction) must therefore be carried out by one classical addition (resp. subtraction) and one conditional subtraction (resp. addition) which should always be performed -- i.e. the effective operation or a dummy one -- for SSCA immunity. Moreover every operation carried out by the coprocessor requires a constant extra software processing $\delta$ to configure the coprocessor. As a result, the cost of field additions/subtractions is not negligible compared to field multiplications. Fig.~\ref{fig_ema} is an electromagnetic radiation measurement during the execution on a smart card of a 192-bit modular multiplication followed by a modular addition. Large amplitude blocks represent the 32-bit crypto-coprocessor activity while those with smaller amplitude are only CPU processing. In this case the time ratio between modular multiplication and modular addition is approximately $0.3$.

From experiments on different smart cards provided with an arithmetic coprocessor, we estimated the average cost of modular ad\-di\-tions/sub\-trac\-tions compared to modular multiplications. Our results are presented in Table~\ref{tab:comp_am} where $A$ and $M$ denote the cost of a field addition/subtraction and the cost of a field multiplication respectively. We observe that the average value of $A/M$ for considered bit lengths is about $0.2$.

\begin{table}[ht]
\centering
\begin{tabular}{|l|c|c|c|c|c|c|c|c|}
\hline
\,Bit length\, & 160 & 192 & 224 & 256 & 320 & 384 & 512 & 521\\
\hline
\,$A/M$ & \,0.36\, & \,0.30\, & \,0.25\, & \,0.22\, & \,0.16\, & \,0.13\, & \,0.09\, & \,0.09\,\\
\hline
\end{tabular}
\smallskip\\
\caption{Measured $A/M$ ratio on smart cards with crypto-coprocessor for NIST and Brainpool ECC bit lengths.}
\label{tab:comp_am}
\end{table}

\vspace{-6mm}
Another useful field operation is negation in $\F_p$, i.e. the map $x \rightarrow -x$, which can be carried out by one non-modular subtraction $p-x$. The cost $N$ of this operation is therefore half the cost of modular addition/subtraction and thus we fix $N/M=0.5 \, A/M$.

In the following sections we also consider the cost $S$ of field squaring. The cost of a squaring compared to a multiplication depends on the functionalities of the corresponding crypto-coprocessor. When a dedicated squaring is available a commonly accepted value for $S/M$ is $0.8$~\cite{BHLM01,GAST05} which is also corroborated by our experiments. Otherwise squarings must be carried out as multiplications and the ratio $S/M$ is thus $1$.

\section{Elliptic Curves}\label{sec:eca}

In this section we recall some generalities about elliptic curves, and useful point representations. Then we present two efficient scalar multiplication algorithms.

Cryptology makes use of elliptic curves over binary fields $\F_{2^n}$ and large characteristic prime fields $\F_p$. In this study we focus on the latter case and hence assume $p>3$.

\subsection{Group Law Over $\F_p$}

An elliptic curve $\E$ over $\F_p$, $p>3$ can be defined as an algebraic curve of affine Weierstra{\ss} equation:
\begin{equation}\label{eq:CE}
\E: y^2 = x^3 + ax + b
\end{equation}
where $a, b \in \F_p$ and $4a^3+27b^2 \not\equiv 0\pmod{p}$.

The set of points of $\E$ -- i.e. the pairs $(x,y) \in {\F_p}^2$ satisfying \eqref{eq:CE} --, plus an extra point $\O$ called \emph{point at infinity} form an abelian group where $\O$ is the neutral element. In the following, we present the corresponding law depending on the selected point representation.

\subsubsection{Affine Coordinates.}

Under the group law a point $P=\left(x_1, y_1\right)$ lying on the elliptic curve $\E$ admits an opposite $-P=\left(x_1, -y_1\right)$.

The sum of $P = \left(x_1, y_1\right)$ and $Q = \left(x_2, y_2\right)$, with $P, Q \ne \mathcal{O}$ and $P \ne \pm Q$, is the point $P+Q=\left(x_3, y_3\right)$ such that:

\begin{equation}
\left\{\begin{array}{rcl}
x_3 &=& ((y_2-y_1)/(x_2-x_1))^2 - x_1 - x_2\\
y_3 &=& (x_1-x_3) (y_2-y_1)/(x_2-x_1) - y_1
\end{array}\right.
\end{equation}

The double of the point $P = \left(x_1, y_1\right)$, with $P \ne \mathcal{O}$ and $y_1 \ne 0$, is the point $2P=\left(x_2, y_2\right)$ as defined below, or $\mathcal{O}$ if $y_1 = 0$.

\begin{equation}
\left\{\begin{array}{rcl}
x_2 &=& ((3{x_1}^2+a)/(2y_1))^2 - 2x_1\\
y_2 &=& (x_1-x_2) (3{x_1}^2+a)/(2y_1) - y_1
\end{array}\right.
\end{equation}

Each point addition or point doubling requires an inversion in $\F_p$. This operation can be very time consuming and leads developers on embedded devices to use other kinds of representations with which point operations involve no field inversion. In the following part of this section, we detail two of them.

\subsubsection{Jacobian Projective Coordinates.}
\label{representation}

By denoting $x = X/Z^2$ and $y = Y/Z^3$, $Z \ne 0$, we obtain the Jacobian projective Weierstra{\ss} equation of the elliptic curve $\E$:
\begin{equation}\label{eq:CE-jacobienne}
Y^2 = X^3+aXZ^4+bZ^6\enspace,
\end{equation}
where $a,b\in\F_p$ and $4a^3+27b^2\neq 0$. Each point $P = (x, y)$ can be represented by its Jacobian projective coordinates $(q^2x: q^3y: q)$ with $q \in \F_p$. Conversely, every point $P = (X: Y: Z)$ different from $\mathcal{O}$ can be represented in affine coordinates by $(x, y) = (X/Z^2, Y/Z^3)$.% Such a curve admits a unique point with $Z = 0$: the point at infinity $\mathcal{O}$ represented by $(q^2: q^3: 0)$ with $q \in \K^*$.

The opposite of a point $\left(X:Y:Z\right)$ is $\left(X:-Y:Z\right)$ and the point at infinity $\mathcal{O}$ is denoted by the unique point with $Z=0$,  $\mathcal{O}=\left(1:1:0\right)$.

The sum of $P = \left(X_1:Y_1:Z_1\right)$ and $Q = \left(X_2:Y_2:Z_2\right)$, with $P, Q \ne \mathcal{O}$ and $P \ne \pm Q$, is the point $P+Q=\left(X_3:Y_3:Z_3\right)$ such that:
\begin{equation}\label{eq:jac-proj-add}
\left \{
\begin{array}{c @{\;=\;} l}
X_3 & F^2-E^3-2AE^2\\
Y_3 & F\left(AE^2-X_3\right)-CE^3\\
Z_3 & Z_1Z_2E\\
\end{array}
\right.
\text{\quad with \quad}
\begin{array}{c @{\;=\;} l}
A & X_1{Z_2}^2\\
B & X_2{Z_1}^2\\
C & Y_1{Z_2}^3\\
D & Y_2{Z_1}^3\\
E & B-A\\
F & D-C\\
\end{array}
\end{equation}

If $P$ is given in affine coordinates -- i.e. $Z_1=1$ -- it is possible to save up one field squaring and four multiplications in \eqref{eq:jac-proj-add}. Such a case is referred to as \emph{mixed affine-Jacobian addition}. On the other hand if $P$ has to be added several times, storing ${Z_1}^2$ and ${Z_1}^3$ saves one squaring and one multiplication in all following additions involving $P$. This latter case is referred to as \emph{readdition}.

\pagebreak

The double of the point $P = \left(X_1:Y_1:Z_1\right)$ is the point $2P=\left(X_2:Y_2:Z_2\right)$ such that:

\begin{equation}\label{eq:jac-proj-dbl}
\left \{
\begin{array}{c @{\;=\;} l}
X_2 & C^2-2B\\
Y_2 & C\left(B-X_2\right)-2A^2\\
Z_2 & 2Y_1Z_1\\
\end{array}
\right.
\text{\quad with \quad}
\begin{array}{c @{\;=\;} l}
A & 2{Y_1}^2\\
B & 2AX_1\\
C & 3{X_1}^2+a{Z_1}^4\\
\end{array}\end{equation}

When curve parameter $a$ is $-3$, doubling can be carried out taking $C = 3 \left( X_1+{Z_1}^2 \right) \left( X_1-{Z_1}^2 \right)$ which saves two squarings in \eqref{eq:jac-proj-dbl}. We denote this operation by \emph{fast doubling}.

Adding up field operations yields $12M+4S+7A$ for general addition, $11M+3S+7A$ for readdition, $8M+3S+7A$ for mixed addition, $4M+6S+11A$ for general doubling formula and $4M+4S+12A$ for fast doubling.

\subsubsection{Modified Jacobian Projective Coordinates.}
\label{repr_jac_mod}

This representation, introduced in~\cite{COM98}, is derived from the Jacobian projective representation to which a fourth coordinate is added for computation convenience. In this representation, a point on the curve $\E$ is thus represented by $(X : Y : Z : aZ^4)$, where $(X:Y:Z)$ stands for the Jacobian representation.

Modified Jacobian projective coordinates provide a particularly efficient doubling formula. Indeed, the double of a point $P = \left(X_1:Y_1:Z_1:W_1\right)$ is given by $2P=\left(X_2:Y_2:Z_2:W_2\right)$ such that:

\begin{equation}\label{eq:modified-jac-proj-dbl}
\left \{
\begin{array}{c @{\;=\;} l}
X_2 & A^2 - 2C\\
Y_2 & A\left(C-X_2\right)-D\\
Z_2 & 2Y_1Z_1\\
W_2 & 2DW_1\\
\end{array}
\right.
\text{\quad with \quad}
\begin{array}{c @{\;=\;} l}
A & 3{X_1}^2+W_1\\
B & 2{Y_1}^2\\
C & 2B{X_1}\\
D & 2B^2\\
\end{array}
\end{equation}

Doubling hence requires only $4M+4S+12A$ for all $a$ values. On the other hand, addition is less efficient compared to Jacobian projective representation: by applying formula~\eqref{eq:jac-proj-add}, we need to compute the fourth coordinate which is required in point doubling, adding an overhead of $1M+2S$~\cite{Joy08a}.

\subsubsection{On S--M Trade-Offs.}

Addition and doubling formulas presented above are voluntarily not state-of-the-art, see~\cite{BL_EFD}. Indeed, recent advances have provided Jacobian formulas where some field multiplications have been traded for faster field squarings~\cite[Sec.~4.1]{Lon07}. These advances have been achieved by using the so-called \emph{S--M trade-off} principle which is based on the fact that computing $ab$ when $a^2$ and $b^2$ are known can be done as $2ab=(a+b)^2-a^2-b^2$. This allows a squaring to replace a multiplication since the additional factor 2 can be handled by considering the representative of the Jacobian coordinates equivalence class $(X:Y:Z)=(2^2X:2^3Y:2Z)$.

Nevertheless such trade-offs not only replace field multiplications by field squarings but also add field additions. In the previous example at least $3$ extra additions have to be performed, thus taking $S/M=0.8$ implies that the trade-off is profitable only if $A/M<0.067$ which is never the case with devices considered using standardized curves as seen in Section~\ref{add_cost}. These new formulas are thus not relevant in the context of embedded devices.

\subsection{Scalar Multiplication}\label{multscal}

\subsubsection{Generalities.}

The operation consisting in calculating the multiple of a point $k\cdot P = P+P+\cdots+P$ ($k$ times) is called \emph{scalar multiplication} and the integer $k$ is thus referred to as the \emph{scalar}.

Scalar multiplication is used in ECDSA signature~\cite{ANSIX9.62-2005} and ECDH key agreement~\cite{ANSIX9.63-2001} protocols. Implementing such protocols on embedded devices requires particular care from both the efficiency and the security points of view. Indeed scalar multiplication turns out to be the most time consuming part of the aforementioned protocols, and since it uses secret values as scalars,  side-channel analysis endangers the security of those protocols.

Most of the scalar multiplication algorithms published so far are derived from the traditional \textit{double and add} algorithm. This algorithm can scan the binary representation of the scalar in both directions which leads to the \textit{left-to-right} and \textit{right-to-left} variants. The former is generally preferred over the latter since it saves one point in memory.

Moreover since computing the opposite of a point $P$ on an elliptic curve is virtually free, the most efficient methods for scalar multiplication use signed digit representations such as the Non-Adjacent Form (NAF)~\cite{AW93}. Under the NAF representation, an $n$-bit scalar has an average Hamming weight of $n/3$ which implies that one point doubling is performed every bit of scalar and one point addition is performed every three bits.

In the two next subsections, we present a left-to-right and a right-to-left NAF scalar multiplication algorithms.

\subsubsection{Left To Right Binary NAF Scalar Multiplication.}
\label{mult_scal_lr}
Alg.~\ref{alg:mult-naf-lr} presents the classical NAF scalar multiplication algorithm.

\begin{algo}
 {Left-to-right binary NAF scalar multiplication~\cite{HMV04}}
 {$P = (X_1 : Y_1 : Z_1) \in \EFp$, $k=(k_{l-1} \dots k_1 k_0)_{\text{NAF}}$}
 {$k\cdot P$}\label{alg:mult-naf-lr}
 \ns{$(X_2 : Y_2 : Z_2) \leftarrow (X_1 : Y_1 : Z_1)$}
 \ns{$i \leftarrow l-2$}
 \ns{\While $i \geq 0$ \Do}
 \nl{$(X_2 : Y_2 : Z_2) \leftarrow 2 \cdot (X_2 : Y_2 : Z_2)$}
 \nl{\indent \If $k_i = 1$ \Then}
 \nl{\hspace*{5mm}\indent\indent $(X_2 : Y_2 : Z_2) \leftarrow (X_2 : Y_2 : Z_2) + (X_1 : Y_1 : Z_1)$}
 \nl{\indent \If $k_i = -1$ \Then}
 \nl{\hspace*{5mm}\indent\indent $(X_2 : Y_2 : Z_2) \leftarrow (X_2 : Y_2 : Z_2) - (X_1 : Y_1 : Z_1)$}
 \nl{$i \leftarrow i-1$}
 \ns{\Ret $(X_2 : Y_2 : Z_2)$}
\end{algo}

Point doubling can be done in Alg. \ref{alg:mult-naf-lr} using general Jacobian doubling formula or fast doubling formula. Since NIST curves fulfill $a=-3$ and each Brainpool curve is provided with an isomorphism to a curve with $a = -3$, we thus assume that fast doubling is always possible. Point addition can be performed using mixed addition formula if input points are given in affine coordinates or by using readdition formula otherwise.

It is possible to reduce the number of point additions by using window techniques\footnote{By \emph{window techniques} we mean the sliding window NAF and the Window NAF$_w$ algorithms, see~\cite{HMV04} for more details.} which need the precomputation of some first odd multiples of the point $P$. Table~\ref{tab:nb_add} recalls the number of point additions per bit of scalar when having from 0 (simple NAF) to 4 precomputed points. More than 4 points allows even better results but seems not practical in the context of constrained memory.

\begin{table}%
\center
\begin{tabular}{|l|c|c|c|c|c|}
\hline
\,Nb. of precomp. points\, & 0 & 1 & 2 & 3 & 4 \\
\hline
\,Precomputed points\, & -- & $3P$ & $3P,5P$ & $3P,5P,7P$ & $3P, \dots, 9P$ \\
\hline
\,Point additions / bit\, & \,$1/3 \approx 0.33$\, & \,$1/4=0.25$\, & \,$2/9 \approx 0.22$\, & \,$1/5=0.20$\, & \,$4/21 \approx 0.19$\, \\
\hline
\end{tabular}
\smallskip\\
\caption{Average number of point additions per bit of scalar using window NAF algorithms.}
\label{tab:nb_add}
\end{table}

\vspace{-10mm}

\subsubsection{Right To Left Binary NAF Mixed Coordinates Multiplication.}
\label{mult_scal_mixt}

We recall here a very efficient algorithm performing right-to-left NAF scalar multiplication. Indeed this algorithm uses the fast modified Jacobian doubling formula which works for all curves -- i.e. for all $a$ -- without needing the slow modified Jacobian addition.

This is achieved by reusing the idea of \emph{mixed coordinates} scalar multiplication (i.e. two coordinate systems are used simultaneously) introduced by Cohen, Ono and Miyaji in~\cite{COM98}. The aim of this approach is to make the best use of two coordinates systems by processing some operations with one system and others with the second. Joye proposed in~\cite{Joy08a} to perform additions by using Jacobian coordinates, doublings -- referred to as $*$ -- by using modified Jacobian coordinates, and to compute the NAF representation of the scalar on-the-fly, cf. Alg.~\ref{alg:mult-naf-mixt}\footnote{In Alg.~\ref{alg:mult-naf-mixt}, Jacobian addition is assumed to handle the special cases $P = \pm Q$, $P=\mathcal{O}$, $Q=\mathcal{O}$ as discussed in~\cite{Joy08a}.}.

%% DEPLACE AVANT L'ALGO POUR MISE EN PAGE
In the same way as their left-to-right counterpart benefits from precomputed points, right-to-left algorithms can be enhanced using window techniques if extra memory is available~\cite{Yao76, Joy09}. In this case precomputations are replaced by postcomputations the cost of which is negligible for the considered window sizes and bit lengths.

In~\cite{Joy08a} the author suggests protecting Alg.~\ref{alg:mult-naf-mixt} against SSCA by using the so-called atomicity principle. We recall in the next section the principle of this SSCA countermeasure.
%% FIN

\begin{algo}
 {Right-to-left binary NAF mixed coordinates multiplication~\cite{Joy08a}}
 {$P = (X_1 : Y_1 : Z_1) \in \EFp$, $k$}
 {$k\cdot P$}\label{alg:mult-naf-mixt}
 \ns{$(X_2 : Y_2 : Z_2) \leftarrow (1:1:0)$}
 \ns{$(R_1 : R_2 : R_3 : R_4) \leftarrow (X_1:Y_1:Z_1:a{Z_1}^4)$}
 \ns{\While $k > 1$ \Do}
 \nl{\indent \If $k \equiv 1 \mod 2$ \Then}
 \nl{\hspace*{6mm}\indent\indent $u\leftarrow 2- (k \mod 4)$}
 \nl{\hspace*{5mm} $k \leftarrow k-u$}
 \nl{\hspace*{5mm} \If $u = 1$ \Then}
 \nl{\hspace*{10mm} $(X_2 : Y_2 : Z_2) \leftarrow (X_2 : Y_2 : Z_2)+(R_1 : R_2 : R_3)$}
 \nl{\hspace*{5mm} \Else}
 \nl{\hspace*{10mm} $(X_2 : Y_2 : Z_2) \leftarrow (X_2 : Y_2 : Z_2)+(R_1 : -R_2 : R_3)$}
 \nl{$k\leftarrow k/2$}
 \nl{$(R_1 : R_2 : R_3 : R_4) \leftarrow 2*(R_1 : R_2 : R_3 : R_4)$}
 \ns{$(X_2 : Y_2 : Z_2) \leftarrow (X_2 : Y_2 : Z_2)+(R_1 : R_2 : R_3)$}
 \ns{\Ret $(X_2 : Y_2 : Z_2)$}
\end{algo}

\section{Atomicity}\label{sec:atomicite}

In this section we recall the principle of atomicity and its application to scalar multiplication. Other countermeasures exist in order to thwart SSCA such as regular algorithms~\cite{Cor99,JY02,Joy09} and unified formulas~\cite{BJ02b,Edw07}. However regular algorithms require costly extra curve operations, and unified formulas for Weierstrass curves over $\F_p$ -- only known in the affine and homogeneous coordinate systems, see~\cite{BJ02b} -- are also very costly. Therefore atomicity turns out to be more efficient in the context of embedded devices. It is thus natural to compare the efficiency of the two scalar multiplication methods presented in Section~\ref{multscal} protected by atomicity.

We recall in the following how atomicity is generally implemented on elliptic curves cryptography, for a complete atomicity principle description see \cite{CCJ04}.

\subsection{State-of-the-Art}

The atomicity principle has been introduced in~\cite{CJ02brevet}. This countermeasure consists in rewriting all the operations carried out through an algorithm into a sequence of identical \emph{atomic patterns}. The purpose of this method is to defeat SSCA since an attacker has nothing to learn from an uniform succession of identical patterns.

In the case of scalar multiplications, a succession of point doublings and point additions is performed. Each of these operations being composed of field operations, the execution of a scalar multiplication can be seen as a succession of field operations. The atomicity consists here in rewriting the succession of field operations into a sequence of identical atomic patterns. The atomic pattern (1) proposed in \cite{CCJ04} is composed of the following field operations: a multiplication, two additions and a negation. $R_i$'s denote the crypto-coprocessor registers.
$$(1)
\left[
\begin{array}{l @{\;\leftarrow\;} l}
R_{1} & R_{2} \cdot R_{3}\\
R_{4} & R_{5} + R_{6}\\
R_{7} & -R_{8}\\
R_{9} & R_{10} + R_{11}\\
\end{array}
\right .$$
This choice relies on the observation that during the execution of point additions and point doublings, no more than two additions and one negation are required between two multiplications. Atomicity consists then of writing point addition and point doubling as sequences of this pattern -- as many as there are field multiplications (including squarings).

Therefore this countermeasure induces two kinds of costs:
\begin{itemize}
	\item Field squarings have to be performed as field multiplications. Then this approach is costly on embedded devices with dedicated hardware offering modular squaring operation, i.e. when $S/M<1$.
	\item Dummy additions and negations are added. Their cost is generally negligible from a theoretical point of view but, as shown in Section~\ref{add_cost}, the cost of such operations must be taken into account in the context of embedded devices.
\end{itemize}

To reduce these costs, Longa proposed in his PhD thesis~\cite[Chap.~5]{Lon07} the two following atomic patterns in the context of Jacobian coordinates:

$$
(2)
\left[
\begin{array}{l @{\;\leftarrow\;} l}
R_{1} & R_{2} \cdot R_{3}\\
R_{4} & -R_{5}\\
R_{6} & R_{7} + R_{8}\\
R_{9} & R_{10} \cdot R_{11}\\
R_{12} & -R_{13}\\
R_{14} & R_{15} + R_{16}\\
R_{17} & R_{18} + R_{19}\\
\end{array}
\right .
\quad
(3)
\left[
\begin{array}{l @{\;\leftarrow\;} l}
R_{1} & {R_{2}}^2\\
R_{3} & -R_{4}\\
R_{5} & R_{6} + R_{7}\\
R_{8} & R_{9} \cdot R_{10}\\
R_{11} & -R_{12}\\
R_{13} & R_{14} + R_{15}\\
R_{16} & R_{17} + R_{18}\\
\end{array}
\right .$$

Compared with atomic pattern (1), these two patterns slightly reduce the number of field additions (gain of one addition every two multiplications). Moreover, atomic pattern (3) takes advantage of the squaring operation by replacing one multiplication out of two by a squaring.

In~\cite[Appendices]{Lon07} Longa expresses mixed affine-Jacobian addition formula as 6 atomic patterns (2) or (3) and fast doubling formula as 4 atomic patterns (2) or (3). It allows to perform an efficient left-to-right scalar multiplication using fast doubling and mixed affine-Jacobian addition protected with atomic patterns (2) or (3).

\subsection{Atomic Left-to-Right Scalar Multiplication}

We detail in the following why the Longa's left-to-right scalar multiplication using fast doubling and mixed affine-Jacobian addition is not compatible with our security constraints.

Defeating DSCA\footnote{We include in DSCA the Template Attack on ECDSA from \cite{MO08}.} requires the randomization of input point coordinates. This can be achieved by two means: projective coordinates randomization~\cite{Cor99} and random curve isomorphism~\cite{JT01}. The first one allows to use the fast point doubling formula but prevents the use of mixed additions since input points $P, 3P, \dots$ have their $Z$ coordinate randomized. On the other hand the random curve isomorphism keeps input points in affine coordinates but randomizes $a$ which thus imposes the use of the general doubling formula instead of the fast one.

Since Longa didn't investigate general doubling nor readdition, we present in Appendix~\ref{appendix:dblgen-mnamnaa} the formulas to perform the former by using 5 atomic patterns (2) or (3) and in Appendix~\ref{appendix:readdition-mnamnaa} the formulas to perform the latter by using 7 atomic patterns (2). It seems very unlikely that one can express readdition using atomic pattern (3): since state-of-the-art readdition formula using the S--M trade-off requires 10 multiplications and 4 squarings, 3 other multiplications would have to be traded for squarings.

Therefore secure left-to-right scalar multiplication can be achieved either by using atomic pattern (2) and projective coordinates randomization which would involve fast doublings and readditions or by using atomic pattern (3) and random curve isomorphism which would involve general doublings and mixed additions.

\subsection{Atomic Right-to-Left Mixed Scalar Multiplication}
\label{sec:atom-rtl}
As suggested in \cite{Joy08a} we protected Alg.~\ref{alg:mult-naf-mixt} with atomicity. Since Longa's atomic patterns have not been designed for modified Jacobian doubling, we applied atomic pattern (1) to protect Alg.~\ref{alg:mult-naf-mixt}.

The decomposition of general Jacobian addition formula in 16 atomic patterns (1) is given in \cite{CCJ04}. Since we haven't found it in the literature, we present in Appendix~\ref{appendix:moddbl-mana} a decomposition of modified Jacobian doubling formula in 8 atomic patterns (1).

Projective coordinates randomization and random curve isomorphism countermeasures can both be applied to this solution.

\subsection{Atomic Scalar Multiplication Algorithms Comparison}
\label{sec:comp1}
We compare in Table~\ref{tab:comp1} the three previously proposed atomically protected algorithms. As discussed in Section~\ref{add_cost} we fix $A/M=0.2$ and $N/M=0.1$. Costs are given as the average number of field multiplications per bit of scalar. Each cost is estimated for devices providing dedicated modular squaring -- i.e. $S/M=0.8$ -- or not -- i.e. $S/M=1$. If extra memory is available, precomputations or postcomputations are respectively used to speed up left-to-right and right-to-left scalar multiplications. The pre/postcomputation cost is here not taken into account but is constant for every row of the chart.

\vspace{10mm}

\begin{table}%
\center
\begin{tabular}{|c|c|c|c|c|}
\hline
\,Nb. of extra\, & \multirow{2}{*}{\,$S/M$\,} & \,Left-to-right\, &  \,Left-to-right\, &  \,Right-to-left\, \\
points & & with (2) & with (3) & with (1) \\
\hline
\multirow{2}{*}{0} & 0.8 & 17.7 & 18.2 & 20.0 \\
\cline{2-5}
 & 1 & 17.7 & 19.6 & 20.0 \\
 \hline
\multirow{2}{*}{1} & 0.8 & 16.1 & 16.9 & 18.0 \\
\cline{2-5}
 & 1 & 16.1 & 18.2 & 18.0 \\
 \hline
\multirow{2}{*}{2} & 0.8 & 15.6 & 16.5 & 17.3 \\
\cline{2-5}
 & 1 & 15.6 & 17.7 & 17.3 \\
 \hline
\multirow{2}{*}{3} & 0.8 & 15.1 & 16.1 & 16.8 \\
\cline{2-5}
 & 1 & 15.1 & 17.4 & 16.8 \\
 \hline
\multirow{2}{*}{4} & 0.8 & 14.9 & 16.0 & 16.6 \\
\cline{2-5}
 & 1 & 14.9 & 17.2 & 16.6 \\
 \hline
\end{tabular}\smallskip\\
\caption{Cost estimation in field multiplications per bit of the 3 atomically protected scalar multiplication algorithms with $A/M=0.2$.}
\label{tab:comp1}
\end{table}

It appears that in our context atomic left-to-right scalar multiplication using atomic pattern (2) with fast doubling and readditions is the fastest solution and is, on average for the 10 rows of Table~\ref{tab:comp1}, 10.5\,\% faster than atomic right-to-left mixed scalar multiplication using atomic pattern (1).

In the next section we present our contribution that aims at minimizing the atomicity cost by optimizing the atomic pattern. Then we apply it on the right-to-left mixed scalar multiplication algorithm since efficient patterns are already known for the two left-to-right variants.

\section{Atomic Pattern Improvement}\label{sec:atomic-improvement}

We propose here a twofold atomicity improvement method: firstly, we take advantage of the fact that a squaring can be faster than a multiplication. Secondly, we reduce the number of additions and negations used in atomic patterns in order to increase the efficiency of scalar multiplication.

\subsection{First Part: Atomic Pattern Extension}\label{idee:sec1}

As explained previously, our first idea is to reduce the efficiency loss due to field squarings turned into multiplications.

\subsubsection{Method Presentation.}

Let $O_1$ and $O_2$ be two atomically written operations (point addition and doubling in our case) such that they require $m$ and $n$ atomic patterns respectively. Let us assume that a sub-operation $o_1$ from the atomic pattern (field multiplication in our case) could sometimes be replaced by another preferred sub-operation $o_2$ (such as field squaring). Let us eventually assume that $O_1$ requires at least $m'$ sub-operations $o_1$ (along with $m-m'$ sub-operations $o_2$) and $O_2$ requires at least $n'$ sub-operations $o_1$ (along with $n-n'$ sub-operations $o_2$).

Then, if $d=\gcd(m, n)>1$, let $e$ represents the greatest positive integer satisfying:
\begin{equation}
e \cdot \dfrac{m}{d}  \leq m-m' \text{\quad and \quad} e \cdot \dfrac{n}{d}  \leq n-n' \enspace .
\end{equation}
Since $0$ is obviously a solution, it is certain that $e$ is defined. If $e>0$ we can now apply the following method. Let a new pattern be defined with $d-e$ original atomic patterns followed by $e$ atomic patterns with $o_2$ replacing $o_1$ -- the order can be modified at convenience.

It is now possible to express operations $O_1$ and $O_2$ with $m/d$ and $n/d$ new patterns respectively. Using the new pattern in $O_1$ (resp. $O_2$) instead of the old one allows replacing $e\cdot m/d$ (resp. $e\cdot n/d$) sub-operations $o_1$ by $o_2$.

\subsubsection{Application to Mixed Coordinates Scalar Multiplication.}
\label{pa_mult_mixt}

Applying this method to Alg.~\ref{alg:mult-naf-mixt} yields the following result: $O_1$ being the Jacobian projective addition, $O_2$ the modified Jacobian projective doubling, $o_1$ the field multiplication and $o_2$ the field squaring, then $m$=16, $m'$=11, $n=8$, $n'=3$, $d=8$ and $e=2$.
Therefore we define a new temporary atomic pattern composed of 8 patterns (1) where 2 multiplications are replaced by squarings. We thus have one fourth of the field multiplications carried out as field squarings. This extended pattern would have to be repeated twice for an addition and once for a doubling.

We applied this new approach in Fig.~\ref{tab:multpatterns1} where atomic general Jacobian addition and modified Jacobian doubling are rewritten in order to take advantage of the squarings. We denote by $\star$ the dummy field additions and negations that must be added to complete atomic patterns.

\begin{figure}[ht]
\begin{center}
\begin{tabular}{cccccc}
Add. 1
&
$\left[
\begin{array}{c @{\;} l}
R_1 & \leftarrow {Z_2}^2\\
\star\\
\textcolor{rouge}{\star}\\
\textcolor{rouge}{\star}\\
R_2 & \leftarrow X_1 \cdot R_1\\
\star\\
\textcolor{rouge}{\star}\\
\textcolor{rouge}{\star}\\
R_1 & \leftarrow R_1 \cdot Z_2\\
\star\\
\textcolor{rouge}{\star}\\
\textcolor{rouge}{\star}\\
R_3 & \leftarrow Y_1 \cdot R_1\\
\star\\
\star\\
\star\\
R_1 & \leftarrow {Z_1}^2\\
\textcolor{rouge}{\star}\\
\star\\
\textcolor{rouge}{\star}\\
R_4 & \leftarrow R_1 \cdot X_2\\
\star\\
R_4 & \leftarrow -R_4\\
R_4 & \leftarrow R_2 + R_4\\
R_1 & \leftarrow Z_1 \cdot R_1\\
\star\\
\star\\
\star\\
R_1 & \leftarrow R_1 \cdot Y_2\\
\star\\
R_1 & \leftarrow -R_1\\
R_1 & \leftarrow R_3 + R_1\\
\end{array}
\right .$
&
\hspace{.2cm}
Add. 2
&
$\left[
\begin{array}{c @{\;} l}
R_6 & \leftarrow {R_4}^2\\
\star\\
\textcolor{rouge}{\star}\\
\textcolor{rouge}{\star}\\
R_5 & \leftarrow Z_1 \cdot Z_2\\
\star\\
\textcolor{rouge}{\star}\\
\textcolor{rouge}{\star}\\
Z_3 & \leftarrow R_5 \cdot R_4\\
\star\\
\textcolor{rouge}{\star}\\
\textcolor{rouge}{\star}\\
R_2 & \leftarrow R_2 \cdot R_6\\
\star\\
R_1 & \leftarrow -R_1\\
\star\\
R_5 & \leftarrow {R_1}^2\\
\textcolor{rouge}{\star}\\
%R_2 & \leftarrow R_2 + R_3\\
R_3 & \leftarrow -R_3\\
\textcolor{rouge}{\star}\\
%R_2 & \leftarrow R_3 + R_2\\
R_4 & \leftarrow R_4 \cdot R_6\\
R_6 & \leftarrow R_5 + R_4\\
R_2 & \leftarrow -R_2\\
R_6 & \leftarrow R_6 + R_2\\
R_3 & \leftarrow R_3 \cdot R_4\\
X_3 & \leftarrow R_2 + R_6\\
\star\\
R_2 & \leftarrow X_3 + R_2\\
R_1 & \leftarrow R_1 \cdot R_2\\
Y_3 & \leftarrow R_3 + R_1\\
\star\\
\star\\
\end{array}
\right .$
&
\hspace{.2cm}
Dbl.
&
$\left[
\begin{array}{c @{\;} l}
R_1 & \leftarrow {X_1}^2\\
R_2 & \leftarrow Y_1 + Y_1\\
\textcolor{rouge}{\star}\\
\textcolor{rouge}{\star}\\
Z_2 & \leftarrow R_2\cdot Z_1\\
R_4 & \leftarrow R_1+R_1\\
\textcolor{rouge}{\star}\\
\textcolor{rouge}{\star}\\
R_3 & \leftarrow R_2\cdot Y_1\\
R_6 & \leftarrow R_3 +R_3\\
\textcolor{rouge}{\star}\\
\textcolor{rouge}{\star}\\
R_2 & \leftarrow R_6\cdot R_3\\
R_1  & \leftarrow R_4 + R_1\\
\star\\
R_1 & \leftarrow R_1 + W_1\\
R_3 & \leftarrow {R_1}^2\\
\textcolor{rouge}{\star}\\
\star\\
\textcolor{rouge}{\star}\\
R_4 & \leftarrow R_6\cdot X_1\\
R_5 &\leftarrow W_1+W_1\\
R_4 & \leftarrow -R_4\\
R_3 & \leftarrow R_3+R_4\\

W_2 & \leftarrow R_2\cdot R_5\\
X_2 & \leftarrow R_3+R_4\\
R_2 & \leftarrow -R_2\\
R_6 & \leftarrow R_4+X_2\\

R_4 & \leftarrow R_6\cdot R_1\\
\star\\
R_4 & \leftarrow -R_4\\
Y_2 & \leftarrow R_4+R_2\\

\end{array}
\right .$
\end{tabular}
\end{center}
\caption{Extended atomic pattern applied to Jacobian projective addition and modified Jacobian projective doubling.}%
\label{tab:multpatterns1}%
\end{figure}

\subsection{Second Part: Atomic Pattern Cleaning-Up}\label{idee:sec2}

In a second step we aim at reducing the number of dummy field operations. In Fig.~\ref{tab:multpatterns1}, we identified by $\textcolor{rouge}{\star}$ the operations that are never used in Add.1, Add.2 and Dbl. These field operations may then be removed saving up 5 field additions and 3 field negations per pattern occurrence.

However, we found out that field operations could be rearranged in order to maximize the number of rows over the three columns composed of dummy operations only. We then merge negations and additions into subtractions when possible. This improvement is depicted in Fig.~\ref{tab:multpatterns2}.

\begin{figure}[ht]
\begin{center}
\begin{tabular}{cccccc}
Add. 1
&
$\left[
\begin{array}{c @{\;} l}
R_1 & \leftarrow {Z_2}^2\\
\star\\
\textcolor{rouge}{\star}\\
\textcolor{rouge}{\star}\\
R_2 & \leftarrow Y_1 \cdot Z_2\\
\star\\
\textcolor{rouge}{\star}\\
\textcolor{rouge}{\star}\\
R_5 & \leftarrow Y_2 \cdot Z_1\\
\star\\
\textcolor{rouge}{\star}\\
\textcolor{rouge}{\star}\\
R_3 & \leftarrow R_1 \cdot R_2\\
\star\\
\textcolor{rouge}{\star}\\
\star\\
R_4 & \leftarrow {Z_1}^2\\
\textcolor{rouge}{\star}\\
\textcolor{rouge}{\star}\\
\textcolor{rouge}{\star}\\
R_2 & \leftarrow R_5\cdot R_4\\
\star\\
\textcolor{rouge}{\star}\\
R_2 & \leftarrow R_2 - R_3\\
R_5 & \leftarrow R_1 \cdot X_1\\
\star\\
\textcolor{rouge}{\star}\\
\star\\
R_6 & \leftarrow X_2 \cdot R_4\\
\textcolor{rouge}{\star}\\
\textcolor{rouge}{\star}\\
R_6 & \leftarrow R_6 - R_5\\
\end{array}
\right .$
&
\hspace{.2cm}
Add. 2
&
$\left[
\begin{array}{c @{\;} l}
R_1 & \leftarrow {R_6}^2\\
\star\\
\textcolor{rouge}{\star}\\
\textcolor{rouge}{\star}\\
R_4 & \leftarrow R_5 \cdot R_1\\
\star\\
\textcolor{rouge}{\star}\\
\textcolor{rouge}{\star}\\
R_5 & \leftarrow R_1 \cdot R_6\\
\star\\
\textcolor{rouge}{\star}\\
\textcolor{rouge}{\star}\\
R_1 & \leftarrow Z_1 \cdot R_6\\
\star\\
\textcolor{rouge}{\star}\\
\star\\
R_6 & \leftarrow {R_2}^2\\
\textcolor{rouge}{\star}\\
\textcolor{rouge}{\star}\\
\textcolor{rouge}{\star}\\
Z_3 & \leftarrow R_1 \cdot Z_2\\
R_1 & \leftarrow R_4+R_4\\
\textcolor{rouge}{\star}\\
R_6 & \leftarrow R_6 - R_1\\
R_1 & \leftarrow R_5 \cdot R_3\\
X_3 & \leftarrow R_6 - R_5\\
\textcolor{rouge}{\star}\\
R_4 & \leftarrow R_4 - X_3\\
R_3 & \leftarrow R_4 \cdot R_2\\
\textcolor{rouge}{\star}\\
\textcolor{rouge}{\star}\\
Y_3 & \leftarrow R_3 - R_1\\
\end{array}
\right .$
&
\hspace{.2cm}
Dbl.
&
$\left[
\begin{array}{c @{\;} l}
R_1 & \leftarrow {X_1}^2\\
R_2 & \leftarrow Y_1 + Y_1\\
\textcolor{rouge}{\star}\\
\textcolor{rouge}{\star}\\
Z_2 & \leftarrow R_2\cdot Z_1\\
R_4 & \leftarrow R_1+R_1\\
\textcolor{rouge}{\star}\\
\textcolor{rouge}{\star}\\
R_3 & \leftarrow R_2\cdot Y_1\\
R_6 & \leftarrow R_3 +R_3\\
\textcolor{rouge}{\star}\\
\textcolor{rouge}{\star}\\
R_2 & \leftarrow R_6\cdot R_3\\
R_1  & \leftarrow R_4 + R_1\\
\textcolor{rouge}{\star}\\
R_1 & \leftarrow R_1 + W_1\\
R_3 & \leftarrow {R_1}^2\\
\textcolor{rouge}{\star}\\
\textcolor{rouge}{\star}\\
\textcolor{rouge}{\star}\\
R_4 & \leftarrow R_6\cdot X_1\\
R_5 &\leftarrow W_1+W_1\\
\textcolor{rouge}{\star}\\
R_3 & \leftarrow R_3-R_4\\

W_2 & \leftarrow R_2\cdot R_5\\
X_2 & \leftarrow R_3-R_4\\
\textcolor{rouge}{\star}\\
R_6 & \leftarrow R_4-X_2\\

R_4 & \leftarrow R_6\cdot R_1\\
\textcolor{rouge}{\star}\\
\textcolor{rouge}{\star}\\
Y_2 & \leftarrow R_4-R_2\\

\end{array}
\right .$
\end{tabular}
\end{center}
\caption{Improved arrangement of field operations in extended atomic pattern from Fig.~\ref{tab:multpatterns1}.}
\label{tab:multpatterns2}%
\end{figure}

\clearpage

This final optimization now allows us to save up 6 field additions and to remove the 8 field negations per pattern occurrence. One may note that no more dummy operation remains in modified Jacobian doubling. We thus believe that our resulting atomic pattern (4) is optimal for this operation:

$$
(4)
\left[
\begin{array}{l @{\;\leftarrow\;} l}
R_{1} & {R_{2}}^2\\
R_{3} & R_{4} + R_{5}\\
R_{6} & R_{7} \cdot R_{8}\\
R_{9} & R_{10} + R_{11}\\
R_{12} & R_{13} \cdot R_{14}\\
R_{15} & R_{16} + R_{17}\\
R_{18} & R_{19} \cdot R_{20}\\
R_{21} & R_{22} + R_{23}\\
R_{24} & R_{25} + R_{26}\\
R_{27} & {R_{28}}^2\\
R_{29} & R_{30} \cdot R_{31}\\
R_{32} & R_{33} + R_{34}\\
R_{35} & R_{36} - R_{37}\\
R_{38} & R_{39} \cdot R_{40}\\
R_{41} & R_{42} - R_{43}\\
R_{44} & R_{45} - R_{46}\\
R_{47} & R_{48} \cdot R_{49}\\
R_{50} & R_{51} - R_{52}\\
\end{array}
\right .$$

\vspace{-6mm}\subsection{Theoretical Gain}
In Table~\ref{tab:comp2} we present the cost of right-to-left mixed scalar multiplication protected with atomic pattern (4). We also draw up in this chart the gains obtained over left-to-right and right-to-left algorithms protected with atomic patterns (2) and (1) respectively.
%\vspace{-5mm}
\enlargethispage{1cm}
\begin{table}[ht]
\center
\begin{tabular}{|c|c|c|c|c|}
\hline
\,Nb. of extra\, & \multirow{2}{*}{\,$S/M$\,} & \,Right-to-left\, &  \,Gain over\, & \,Gain over\, \\
points & & with (4) & \,l.-to-r. with (2)\, & \,r.-to-l. with (1)\, \\
\hline
\multirow{2}{*}{0} & 0.8 & 16.0\,M & 9.6\,\% & 20.0\,\% \\
\cline{2-5}
 & 1 & 16.7\,M & 5.6\,\% & 16.5\,\% \\
 \hline
\multirow{2}{*}{1} & 0.8 & 14.4\,M & 10.6\,\% & 20.0\,\% \\
\cline{2-5}
 & 1 & 15.0\,M & 6.8\,\% & 16.7\,\% \\
 \hline
\multirow{2}{*}{2} & 0.8 & 13.9\,M & 10.9\,\% & 19.7\,\% \\
\cline{2-5}
 & 1 & 14.4\,M & 7.7\,\% & 16.8\,\% \\
 \hline
\multirow{2}{*}{3} & 0.8 & 13.4\,M & 11.3\,\% & 20.2\,\% \\
\cline{2-5}
 & 1 & 14.0\,M & 7.3\,\% & 16.7\,\% \\
 \hline
\multirow{2}{*}{4} & 0.8 & 13.3\,M & 10.7\,\% & 19.9\,\% \\
\cline{2-5}
 & 1 & 13.8\,M & 7.4\,\% & 16.9\,\% \\
 \hline
\end{tabular}\smallskip\\
\caption{Costs estimation in field multiplications per bit of Alg.~\ref{alg:mult-naf-mixt} protected with improved pattern (4) and comparison with two others methods presented in Table~\ref{tab:comp1} assuming $A/M=0.2$.}
\label{tab:comp2}
\end{table}

Due to our new atomic pattern (4), right-to-left mixed scalar multiplication turns out to be the fastest among these solutions in every cases. The average speed-up over pattern (1) is 18.3\,\% and the average gain over left-to-right scalar multiplication protected with atomic pattern (2) is 10.6\,\% if dedicated squaring is available or 7.0\,\% otherwise.

\subsection{Experimental Results}

We have implemented Alg.~\ref{alg:mult-naf-mixt} -- without any window method -- protected with the atomic pattern (1) on one hand and with our improved atomic pattern (4) on the other hand. We used a chip equipped with an 8-bit CPU running at 30 MHz and with a 32-bit crypto-coprocessor running at 50 MHz. In particular, this crypto-coprocessor provides a dedicated modular squaring.  The characteristics of the corresponding implementation are given in Table~\ref{tab:carac}. On the NIST P-192 curve~\cite{FIPS186-3} we obtained a practical speed-up of about 14.5\,\% to be compared to the predicted 20\,\%.
This difference can be explained by the extra software processing required in the scalar multiplication loop management, especially the on-the-fly NAF decomposition of the scalar in an SSCA-resistant way.
\begin{table}%
\center
\begin{tabular}{|c|c|c|c|c|}
\hline
\,Timing\, & \,RAM size\, & \,Code size\,  \\
\hline
\,29.6 ms\, & 412 B &  3.5 KB  \\
\hline
\end{tabular}\smallskip\\
\caption{Characteristics of our implementation of the atomically protected 192-bit scalar multiplication on an 8-bit chip with a 32-bit crypto-coprocessor.}
\label{tab:carac}
\end{table}

\begin{figure}[ht]%
\center
\ifpdf
\includegraphics[width=.8\columnwidth]{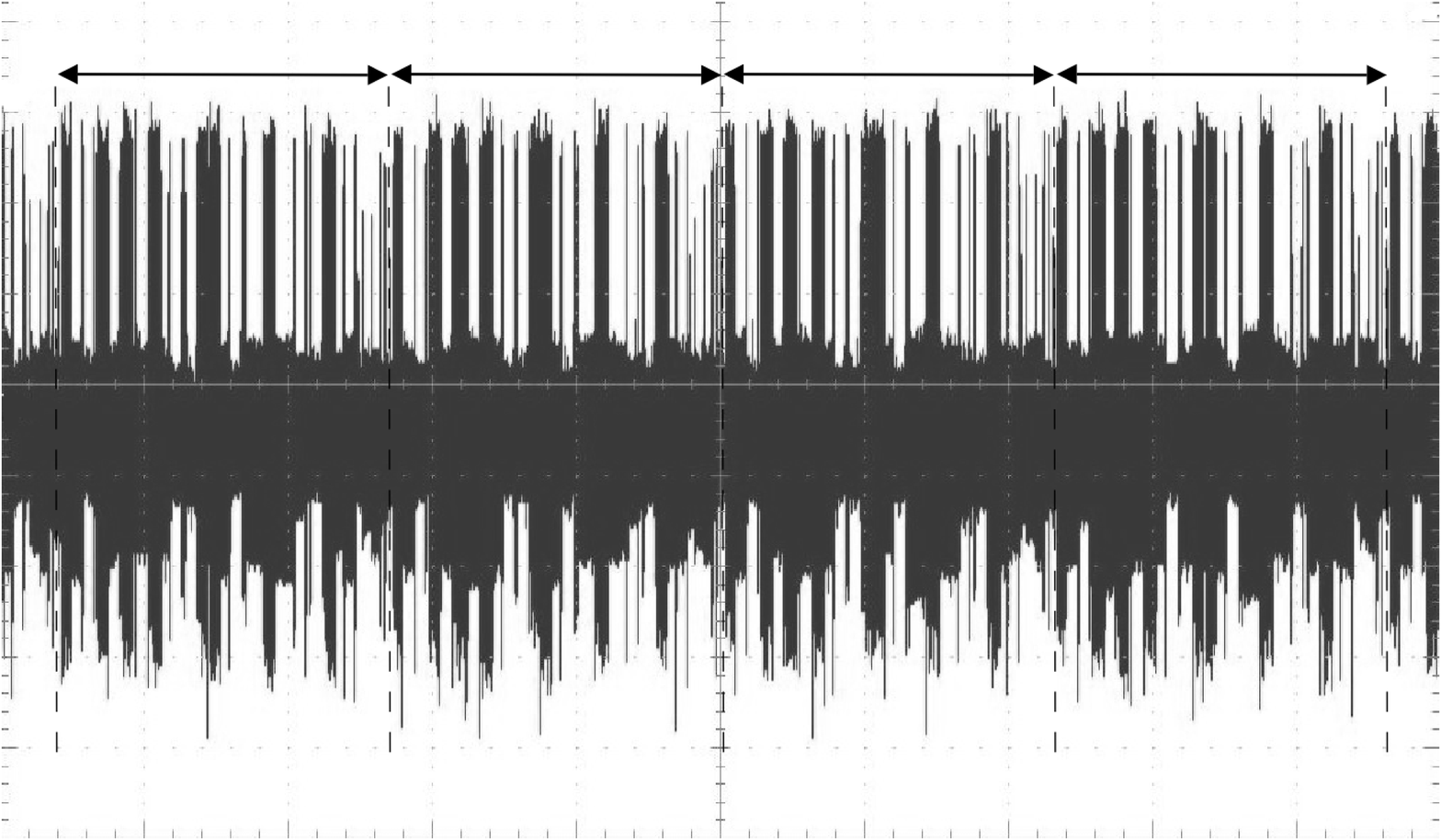}%
\else
\includegraphics[bb=0 0 1400 816,width=.75\columnwidth]{patterns.eps}%
\fi 
\caption{Side-channel leakage observed during the execution of our scalar multiplication implementation showing a sequence of atomic patterns.}%
\label{fig_ema2}%
\end{figure}

\vspace{-9mm} When observing the side-channel leakage of our implementation we obtained the signal presented in Fig.~\ref{fig_ema2}. Atomic patterns comprising 8 modular multiplications and several additions/subtractions can easily be identified.

\enlargethispage{1cm}

\clearpage

\section{Conclusion}\label{sec:conclusion}

In this paper, we propose a new atomic pattern for scalar multiplication on elliptic curves over $\F_p$ and detail our method for atomic pattern improvement.
To achieve this goal, two ways are explored. Firstly we maximize the use of squarings to replace multiplications since the latter are slower. Secondly we minimize the use of field additions and negations since they induce a non-negligible penalty.
In particular, we point out that the classical hypothesis taken by scalar multiplication designers to neglect the cost of additions/subtractions in $\F_p$ is not valid when focusing on embedded devices such as smart cards.

In this context our method provides an average 18.3\,\% improvement for the right-to-left mixed scalar multiplication from~\cite{Joy08a} protected with the atomic pattern from~\cite{CCJ04}. It also provides an average 10.6\,\% gain over the fastest algorithm identified before our contribution if dedicated squaring is available. Furthermore, though the topic of this paper is right-to-left scalar multiplication, our atomic pattern improvement method can be generically used to speed up atomically protected algorithms.

In conclusion we recommend that algorithm designers, addressing the scope of embedded devices, take into account additions and subtractions cost when these operations are heavily used in an algorithm. Moreover the issue of designing efficient atomic patterns should be considered when proposing non regular sensitive algorithms.

\section*{Acknowledgments}

The authors are very grateful to Yannick Sierra for pointing them out the improvement using the subtractions leading to the efficient patterns of Fig.~\ref{tab:multpatterns2}. The authors would also like to thank Christophe Clavier, Sean Commercial, Emmanuelle Dottax, Emmanuel Prouff, Matthieu Rivain and the anonymous referees of Cardis 2010 for their helpful comments on the preliminary versions of this paper.

%\bibliographystyle{abbrv-ocs}
%\bibliography{OCS}

\newpage
\appendix

\section{Atomic Formulas}\label{appendix:atomicformulae}

\subsection{Atomic General Doubling Using Pattern (2) or (3)}
\label{appendix:dblgen-mnamnaa}

The decomposition of a general -- i.e. for all $a$ -- doubling in Jacobian coordinates using atomic pattern (3) is depicted hereafter. The corresponding decomposition using atomic pattern (2) can straightforwardly be obtained by replacing every squaring by a multiplication using the same operand twice.

The input point is given as $(X_1, Y_1, Z_1)$ and the result is written into the point $(X_2, Y_2, Z_2)$. Four intermediate registers, $R_1$ to $R_4$, are used.

\begin{center}
\begin{tabular}{r@{}lr@{}l}
1&$\left[
\begin{array}{c @{\;} l}
R_1 & \leftarrow {X_1}^2\\
\star\\
R_3 & \leftarrow R_1 + R_1\\
R_2 & \leftarrow Z_1 \cdot Z_1\\
\star\\
R_1 & \leftarrow R_1 + R_3\\
R_4 & \leftarrow X_1 + X_1\\
\end{array}
\right .$
&
\hspace{1cm}4&$\left[
\begin{array}{c @{\;} l}
R_2 & \leftarrow {R_1}^2\\
\star\\
\star\\
R_4 & \leftarrow R_4 \cdot R_3\\
R_4 & \leftarrow -R_4\\
R_2 & \leftarrow R_2 + R_4\\
X_2 & \leftarrow R_2 + R_4\\
\end{array}
\right .$
\smallskip\\
2&$\left[
\begin{array}{c @{\;} l}
R_2 & \leftarrow {R_2}^2\\
\star\\
\star\\
R_2 & \leftarrow a \cdot R_2\\
\star\\
R_1 & \leftarrow R_1 + R_2\\
R_2 & \leftarrow Y_1 + Y_1\\
\end{array}
\right .$
&
5& $\left[
\begin{array}{c @{\;} l}
R_3 & \leftarrow {R_3}^2\\
R_1 & \leftarrow -R_1\\
R_4 & \leftarrow X_2 + R_4\\
R_1 & \leftarrow R_1 \cdot R_4\\
R_3 & \leftarrow -R_3\\
R_3 & \leftarrow R_3 + R_3\\
Y_2 & \leftarrow R_3 + R_1\\
\end{array}
\right .$
\smallskip\\
3&$\left[
\begin{array}{c @{\;} l}
R_3 & \leftarrow {Y_1}^2\\
\star\\
\star\\
Z_2 & \leftarrow Z_1 \cdot R_2\\
\star\\
R_3 & \leftarrow R_3 + R_3\\
\star\\
\end{array}
\right .$
&
\end{tabular}\bigskip\\
\end{center}

\newpage
\subsection{Atomic Readdition Using Pattern (2)}
\label{appendix:readdition-mnamnaa}

The decomposition of a readdition in Jacobian coordinates using atomic pattern (2) is depicted hereafter.

The input points are given as $(X_1, Y_1, Z_1)$, $(X_2, Y_2, Z_2)$ and the result is written into the point $(X_3, Y_3, Z_3)$. Seven intermediate registers, $R_1$ to $R_7$, are used.

\begin{center}
\begin{tabular}{r@{}lr@{}l}
1&$\left[
\begin{array}{c @{\;} l}
R_1 & \leftarrow {Y_2}\cdot{Z_1}^3\\
\star\\
\star\\
R_2 & \leftarrow Y_1 \cdot Z_2\\
\star\\
\star\\
\star\\
\end{array}
\right .$
&
\hspace{1cm}5&$\left[
\begin{array}{c @{\;} l}
R_5 & \leftarrow R_5 \cdot R_3\\
\star\\
\star\\
R_3 & \leftarrow Z_2 \cdot R_3\\
\star\\
\star\\
\star\\
\end{array}
\right .$
\smallskip\\
2&$\left[
\begin{array}{c @{\;} l}
R_3 & \leftarrow Z_2 \cdot Z_2\\
\star\\
\star\\
R_4 & \leftarrow R_2 \cdot R_3\\
R_5 & \leftarrow -R_1\\
R_4 & \leftarrow R_4 + R_5\\
\star\\
\end{array}
\right .$
&
6& $\left[
\begin{array}{c @{\;} l}
R_7 & \leftarrow R_4 \cdot R_4\\
\star\\
\star\\
Z_3 & \leftarrow R_3 \cdot Z_1\\
R_5 & \leftarrow -R_5\\
R_7 & \leftarrow R_7 + R_6\\
X_3 & \leftarrow R_7 + R_5\\
\end{array}
\right .$
\smallskip\\
3&$\left[
\begin{array}{c @{\;} l}
R_2 & \leftarrow X_2 \cdot {Z_1}^2\\
\star\\
\star\\
R_3 & \leftarrow X_1 \cdot R_3\\
R_5 & \leftarrow -R_2\\
R_3 & \leftarrow R_3 + R_5\\
\star\\
\end{array}
\right .$
&
7& $\left[
\begin{array}{c @{\;} l}
R_3 & \leftarrow R_1 \cdot R_5\\
R_1 & \leftarrow -X_3\\
R_2 & \leftarrow R_2 + R_1\\
R_1 & \leftarrow R_2 \cdot R_4\\
\star\\
Y_3 & \leftarrow R_1 + R_3\\
\star\\
\end{array}
\right .$
\smallskip\\
4&$\left[
\begin{array}{c @{\;} l}
R_5 & \leftarrow R_3 \cdot R_3\\
\star\\
\star\\
R_2 & \leftarrow R_2 \cdot R_5\\
R_6 & \leftarrow -R_2\\
R_6 & \leftarrow R_6 + R_6\\
\star\\
\end{array}
\right .$
&
\end{tabular}\bigskip\\
\end{center}

\newpage
\subsection{Atomic Modified Jacobian Coordinates Doubling Using Pattern (1)}
\label{appendix:moddbl-mana}

The decomposition of a doubling in modified Jacobian coordinates using atomic pattern (1) is depicted hereafter.

The input point is given as $(X_1, Y_1, Z_1)$ and the result is written into the point $(X_2, Y_2, Z_2)$. Six intermediate registers, $R_1$ to $R_6$, are used.

\begin{center}
\begin{tabular}{r@{}lr@{}l}
1&$\left[
\begin{array}{c @{\;} l}
R_1 & \leftarrow X_1 \cdot X_1\\
R_2 & \leftarrow Y_1 + Y_1\\
\star\\
\star\\
\end{array}
\right .$
&
\hspace{1cm}5&$\left[
\begin{array}{c @{\;} l}
R_3 & \leftarrow R_1 \cdot R_1\\
\star\\
\star\\
\star\\
\end{array}
\right .$
\smallskip\\
2&$\left[
\begin{array}{c @{\;} l}
Z_2 & \leftarrow R_2 \cdot Z_1\\
R_4 & \leftarrow R_1 + R_1\\
\star\\
\star\\
\end{array}
\right .$
&
6&$\left[
\begin{array}{c @{\;} l}
R_4 & \leftarrow R_6 \cdot X_1\\
R_5 & \leftarrow W_1 + W_1\\
R_4 & \leftarrow -R_4\\
R_3 & \leftarrow R_3 + R_4\\
\end{array}
\right .$
\smallskip\\
3&$\left[
\begin{array}{c @{\;} l}
R_3 & \leftarrow R_2 \cdot Y_1\\
R_6 & \leftarrow R_3 + R_3\\
\star\\
\star\\
\end{array}
\right .$
&
7&$\left[
\begin{array}{c @{\;} l}
W_2 & \leftarrow R_2 \cdot R_5\\
X_2 & \leftarrow R_3 + R_4\\
R_2 & \leftarrow -R_2\\
R_6 & \leftarrow R_4 + X_2\\
\end{array}
\right .$
\smallskip\\
4& $\left[
\begin{array}{c @{\;} l}
R_2 & \leftarrow R_6 \cdot R_3\\
R_1 & \leftarrow R_4 + R_1\\
\star\\
R_1 & \leftarrow R_1 + W_1\\
\end{array}
\right .$
&
8&$\left[
\begin{array}{c @{\;} l}
R_4 & \leftarrow R_6 \cdot R_1\\
\star\\
R_4 & \leftarrow -R_4\\
Y_2 & \leftarrow R_4 + R_2\\
\end{array}
\right .$
\end{tabular}\bigskip\\
\end{center}

\end{document}